\documentstyle[prc,aps,epsfig]{revtex}
\tightenlines
\newcommand{\bm}{\bibitem}

\def\be {\begin{equation}}
\def\ee {\end{equation}}
\def\bea {\begin{eqnarray}}
\def\eea {\end{eqnarray}}

\def\ks {{|{\vec k}|}^\ast}

\def\mn {m^{*}_n}
\def\mnsq {m^{*2}_n}

\def\2l {\frac{{f_i}}{(2\lambda + 1)}}
\def\ks {k\!\!\!/}

\def\nn {\nonumber}

\begin{document}
\title{The effects of meson mixing on dilepton spectra}
\author{ O. Teodorescu, A. K. Dutt-Mazumder and C. Gale  }
\address{ Physics Department, McGill University\\ 3600 University St., 
Montreal, Quebec H3A 2T8, Canada\\}
\maketitle

\begin{abstract}
The effect of scalar and vector meson
mixing on the dilepton radiation from hot
and dense hadronic matter is estimated in different isospin channels. 
In particular, we study the effect of $\sigma$-$\omega$ and 
$\rho-a_0$ mixing and calculate the 
corresponding rates. Effects are found to be significant compared to 
standard $\pi$-$\pi$ and $K$-${\bar K}$ annihilations. While the mixing in
the isoscalar channel mostly gives a contribution in the invariant mass range 
between the two-pion 
threshold and the $\omega$ peak, the
isovector channel mixing induces an additional peak just below that of 
the $\phi$.
Experimentally, the dilepton  signals from $\rho$-$a_0$ mixing seem to be more
tractable than those from $\sigma$-$\omega$ mixing.
\end{abstract}
\vspace{0.3 cm}


\vspace{.2 cm}
\section{Introduction}
There exists physical processes which
are forbidden in free space but can take place in matter.  
Those are basically related to medium-induced symmetry breaking effects.  
The interaction 
respects the symmetry, but the ground state breaks it 
\cite{abheemix,broni98,broni00}.
For instance, in matter, Lorentz symmetry is lost which leads to the mixing 
of different spin states even when the interaction Lagrangian respects
all the required symmetry properties.  
A well-known example 
of this is $\sigma$-$\omega$ mixing in nuclear matter 
as was first shown by Chin \cite{chin77} or $\rho$-$a_0$ mixing as
pointed out in Ref.\cite{teodorescu00}. 
In addition there might be mixing between different isospin states
in asymmetric nuclear matter. The $\rho$-$\omega$ mixing   
provides one such example as suggested in  \cite{abheemix}. 

Our goal here is to identify the effects of scalar
and vector meson mixing on dilepton production rates,  which could in
turn be observed
in high energy heavy ion experiments. It
is well known that the electromagnetic radiation provides a penetrating probe 
to study
various in-medium properties of vector mesons. Stimulated by the idea that
in nuclear matter the vector meson properties might be modified from their 
vacuum values as a 
precursor phenomenon to chiral symmetry restoration, a great deal of effort
has been directed to understand their properties in hot and/or dense 
nuclear matter \cite{brown91,rw99}. Several experiments have measured, or 
are planning to 
measure, dilepton pair production in nucleus-nucleus collisions. Such 
measurements have been
carried out by the Dilepton Spectrometer (DLS) at Lawrence Berkeley 
National Laboratory
\cite{dls}, and by HELIOS \cite{hel} and CERES \cite{ceres} at CERN. The NA50 
collaboration also measures dileptons yields in the light vector meson
mass range \cite{NA50}. 
Two new initiatives to 
be mentioned in this context are PHENIX \cite{phenix} at the Relativistic 
Heavy Ion Collider (RHIC) at Brookhaven National Laboratory  and HADES at the 
Gesellschaft f\"ur Schwerionenforschung (GSI)\cite{hades}. 
Density-dependent effects can also be found in experiments
at the Thomas Jefferson National Accelerator Facility (TJNAF) \cite{tjnaf}. As 
the effects to be discussed here are mostly density-driven, the 
latter two facilities represent the energy regime and physical conditions 
perhaps more relevant for the present study. However, the
physics discussed here could also be highlighted in an eventual
low-energy RHIC run.

For the understanding of the data in these experiments
it is important to 
uncover the various processes which might lead to the production of dileptons 
and to estimate their relative contribution to the total production rate. 
Some effort has already being put in obtaining some ``standard''
sources related to the  
decay channels of various light mesons and 
processes like $\pi-\pi$ or $K$-${\bar K}$ annihilation 
\cite{gale87,haglin94,raga99}.
But the opening-up of new channels related to scalar-vector 
mixing, which constitutes the
material to be presented in this paper, has only been addressed recently.

The  mixing of mesons considered here includes both the isoscalar 
and isovector channels.
To be more specific, we estimate the rate of dilepton production from 
$\sigma$-$\omega$ and $\rho$-$a_0$ mixing. Even though some attention 
was paid to the former in the context of heavy-ion collisions 
\cite{wolf98,saito98}, the importance of the latter was shown 
only recently in a zero temperature estimate of the
dilepton production cross section \cite{teodorescu00}. 
The studies of the effects of mixing made so far 
were limited to calculations of cross-sections. 

This paper reports on calculations of the thermal 
production rates of dileptons induced by scalar-vector mixing in the
isoscalar and isovector channels. 
Those results are then 
compared against standard $\pi$-$\pi$, $K$-${\bar K}$
annihilations,  at different temperatures and chemical potentials. 
Note that processes involving mixing of different G-parity states
like $\sigma$-$\omega$ or $\rho$-$a_0$
are allowed only in matter which is not
invariant under charge conjugation. Therefore, 
$\pi~+~\pi$~$\rightarrow$~$e^+e^-$ through their coupling to nucleons
via $\sigma$ and $\omega$ or $\pi~+~\eta~\rightarrow~e^+e^-$ mediated
by $\rho$-$a_0$ mixing,
can take place, as will become clear later, only in matter
with finite nucleon chemical potential. Hence  a natural place to look
for such density-driven effects are experiments involving 
relatively low energy where one expects to have a relatively large chemical
potential.

The model used here is Walecka-like with  
the inclusion of the $a_0$ meson and its mixing with the $\rho$ meson. The
role of $a_0$ in the context of nuclear matter calculation has
also been discussed in Ref. \cite{teodorescu00,kubis,lenske,oset}.
To keep the calculation  simple we take nuclear matter to be symmetric
under $p\leftrightarrow n$ but this can easily be generalized to the case 
where nuclear matter has different number of protons and neutrons.

In section II we discuss the general outline of the formalism. Next in section
III we discuss $\sigma$-$\omega$ and $\rho$-$a_0$ mixing. In section IV 
we present our results for various physical cases. We summarize and
conclude in section IV.

\section{Formalism}

The nucleon-meson 
interaction Lagrangian can be written as
\bea
{\cal L}_{int} = g_\sigma {\bar \psi}\phi_\sigma \psi +
          g_{a_0} {\bar \psi}\phi_{a_0,a}\tau^a \psi
          + g_{\omega NN}{\bar{\psi}} \gamma _\mu\psi\omega^\mu
          + \\
 g_{\rho} [{\bar{\psi}} \gamma _\mu \tau^\alpha
 \psi + \frac{\kappa _\rho}{2m_n}{\bar{\psi}}
    \sigma_{\mu\nu}\tau^\alpha \partial ^\nu] \rho^\mu_\alpha\ ,
\eea
where $\psi$, $\phi_\sigma$, $\phi_{a_0}$, $\rho$ and $\omega$ correspond
to nucleon, $\sigma$, $a_0$ , $\rho$ and $\omega$ fields, and $\tau$s are  
Pauli matrices. The values used for the coupling parameters are obtained
from Ref. \cite{mach89}. From now on we use s and v to denote scalar
and vector mesons, {\em i.e.} s = $\sigma$, $a_0$ and v = $\omega$, $\rho$. It is
understood that $\kappa_\omega=0.0$.

The polarization vector through which the scalar meson couples to the
vector meson via the n-n loop is given by
\bea
\Pi_ \mu (q_0,|{\vec q}|) &=& 2i g_s g_v \int \frac{d^4k}{(2\pi)^4}
    \mbox{Tr}\left[G(k) \Gamma_\mu G(k+q)\right] \ ,\label{pim}
\eea
where 2 is an isospin factor and the vertex for v-nn
coupling is:
\bea
\Gamma_\mu=\gamma_\mu - \frac{\kappa_v}{2m_n}\sigma_{\mu\nu}q^\nu\ .
\label{vertex}
\eea
$G(k)$ is the in-medium nucleon propagator
given by \cite{sewal}
\bea G(k) = G_F(k) + G_T(k)
\label{nuclprop}
\eea
with
\bea
G_F(k)=\frac{(\ks+\mn)}{k^2-m_n^{* 2} + i\epsilon}\ ,
\eea
and
\bea
G_T(k)=(\ks+\mn)\frac{i \pi }{E_k}[\frac{\delta (k_0 - E_k)}
{e^{\beta ({E^*_k} - \nu)}+1} 
 +  
\frac{\delta (k_0 + E_k)}{e^{\beta ({E^*_k} + \nu)}+1}]
\eea
where $E^*_k=\sqrt{{\bf k^2}+m_n^{* 2}}$ and 
$\nu=\mu - \frac{g_v^2 \rho_B}{m_v^2}$ is the effective chemical potential. 
In Eq.~(\ref{nuclprop}), the subscripts
 $F$ and $T$ refer to the free and temperature-dependent part of the 
propagator. $G_F$ is the free Dirac propagator with the  nucleon mass
replaced by its effective value while the 
second term ($G_T$) includes the Fermi distribution of the 
nucleon assuming thermal equilibrium at temperature T. The delta function
guarantees that on-mass-shell levels are occupied
by the appropriate distribution functions. 
 $m_n^*$ denotes the effective nucleon mass evaluated at the mean field 
level \cite{sewal}. It's value is given by the equation:
\bea
\mn=m_n-\frac{g_\sigma^2}{m_\sigma^2}\frac{\gamma}{(2\pi)^3}\int d^3 k\frac{\mn}
{({\bf k^2}+\mnsq)^{1/2}}
[n_k(T)+{\bar n}_k(T)]
\label{mns}
\eea  
where $n_k(T)=\frac{1}{e^{\beta (E^*_k-\nu)}+1}$  ,\hspace{2mm} ${\bar n}_k(T)=\frac{1}{e^{\beta (E^*_k+\nu)}+1}$
are the nucleon and anti-nucleon thermal distributions. 
It should be mentioned
that in asymmetric nuclear matter the mean field generated by the $a_0$ meson would modify the 
neutron and proton mass differently. In the present case this effect 
vanishes. 

In this notation the nucleon density is written as 
\bea
\rho_B=\frac{\gamma}{(2\pi)^3}\int d^3k [n_k(T)-{\bar n}_k(T)]\ .
\label{bardens}
\eea
Now we have a set of two coupled equations, Eqs.~(\ref{mns} and 
\ref{bardens}) and two unknowns: the effective nucleon mass $m_n^*$ 
and the effective chemical potential $\nu$. One looks
for the self-consistent solutions of these two coupled equations in order to
find $m_n^*$ and $\nu$. The results are shown in Fig. 1 for two 
different nuclear densities. At a fixed density the
effective nucleon mass increases and then again drops with temperature (T).
This is consistent with Ref. \cite{sewal}. It might be mentioned here that
we do not include resonances in the present work.

\begin{figure} [htb!]
\begin{center}
\epsfig{file=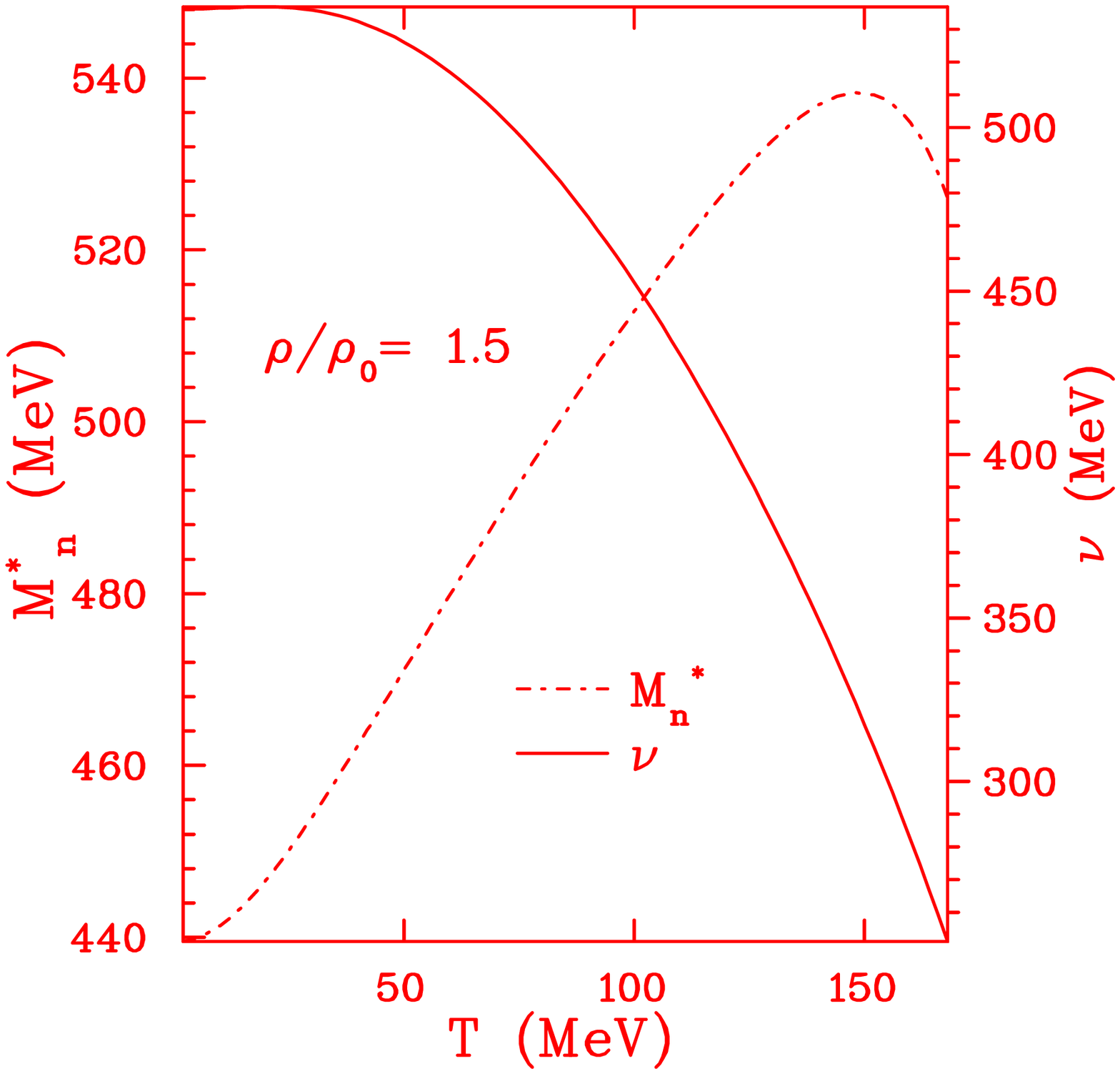,height=5cm,angle=0} 
\epsfig{file=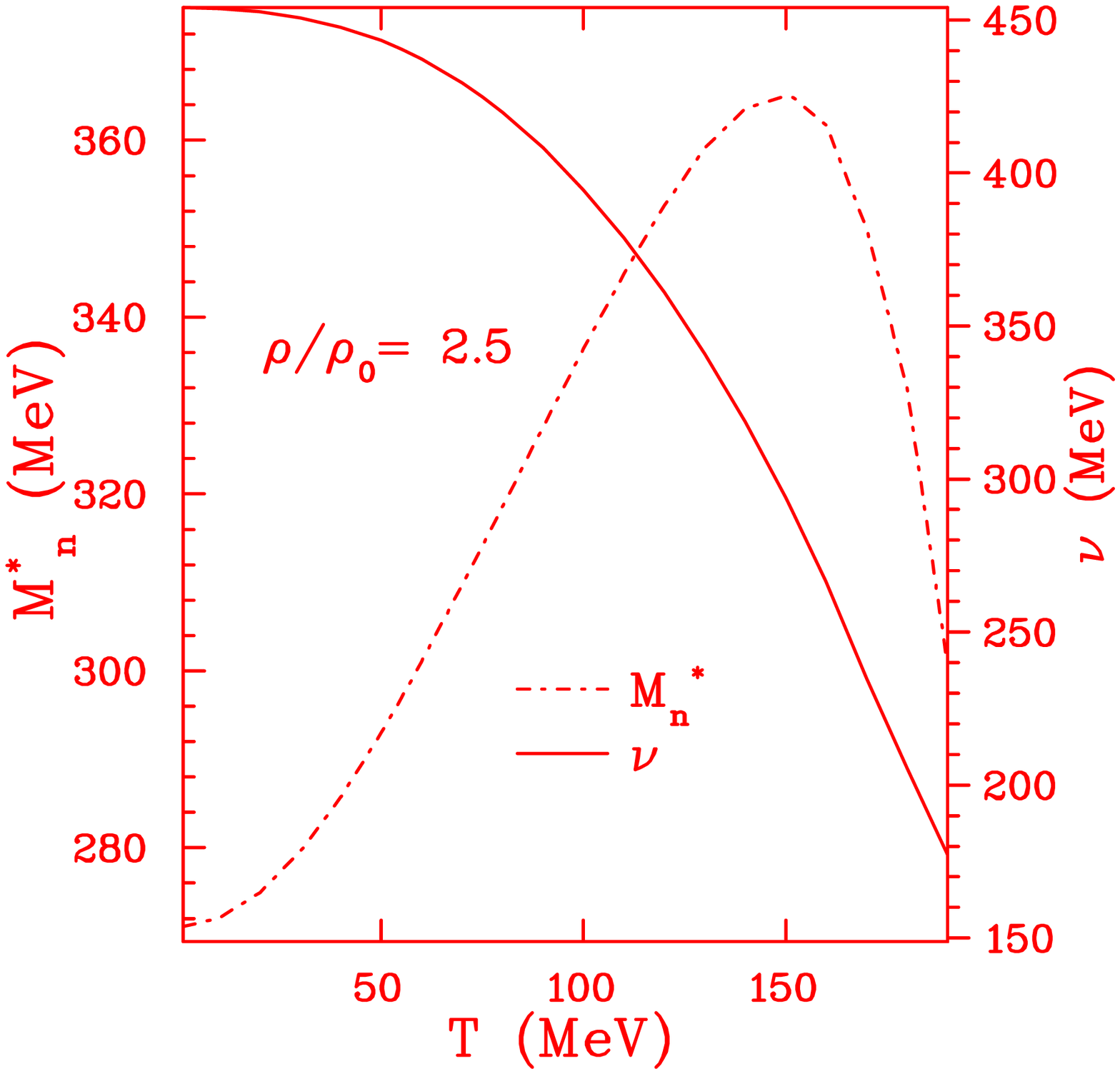,height=5cm,angle=0}
\end{center}
 \caption[Nucleon Mass] 
  {\small The in-medium nucleon mass and chemical potential \\
as function of temperature for two baryonic densities.\label{fig:mnstar}}
\end{figure}

Even though Eq.~(\ref{pim}) involves Green's functions having both vacuum
and temperature dependent part, the vacuum piece for $\Pi_\mu$, as expected 
vanishes on account of Lorentz symmetry. Hence only the thermal part 
contributes to the mixed-polarization function and it has a natural
cut-off provided by the thermal distribution functions. So, 
interestingly, the diagrams involving mixing of different spin states in matter
is free from ultra-violet divergences. 
It should also be
noted that the argument of $\Pi_\mu$ involves 
$q_0$ and $|q|$ indicating Lorentz
symmetry breaking in matter. In general, it could be a function of $\vec q$
but for an isotropic medium it can only depend on the magnitude of the 
three-momentum. 

Returning to Eq.~(\ref{pim}), after the evaluation of the trace and a little 
algebra, $\Pi_\mu$ could be cast into a suggestive form:
\begin{center}
\bea
\Pi_\mu(q_0,|q|)&=&\frac{g_s g_v}{\pi^3} 2q^2 (2m_n^*-
\frac{\kappa q^2}{2 m_n})
\int_0^{\infty}\frac{d^3k}{E^*_k} \sum_{k_0=\pm E_k^*} 
\frac{k_\mu - \frac{q_\mu}{q^2}(k\cdot q)}{q^4 - 4 (k\cdot q)^2} 
\frac{1}{e^{\beta (E_k^* + sgn(k_0) \nu)} +1}\ .
\label{mixamp}
\eea
\end{center}
This immediately leads to two conclusions. First,
it respects the current conservation condition 
$q^\mu\Pi_\mu=0=\Pi_\nu q^\nu$. Secondly, there are only two components which
survive after the integration over azimuthal angle. In fact this
guarantees that it is only the longitudinal component of the $\rho$ meson
which couples to the scalar meson while the transverse
mode remains intact. Furthermore, current conservation implies that
out of the two non-zero components of $\Pi_\mu$, only one is independent.
After performing the angular integration $\Pi_0$ takes the form:
\bea
\Pi_0=\frac{g_s g_v}{8\pi^2 |{\vec q}|}(2m_n^*-\frac{\kappa q^2}{2 m_n}) \int_0^{\infty} \frac{k dk}{E^*_k}
&\{  &2E^*_k\ln[\frac{(q^2+2k|{\vec q}|)^2-4E_k^{*2}E_q^{*2}}
{(q^2-2k|{\vec q}|)^2
-4E_k^{*2}E_q^{*2}}]-q_0\ln[\frac{q^4-4(E^*_kE^*_q-4k|{\vec q}|)^2}
{q^4-4(E^*_k E^*_q+4k|{\vec q}|)^2}] \}  \nn\\
\times &[&\frac{1}{e^{\beta(E_k^*-\nu)}+1}-\frac{1}{e^{\beta(E_k^*+\nu)}+1}]\
.
\label{pi0}
\eea

From Eq.~(\ref{pi0}) it is clear that the mixing amplitude vanishes for 
chemical potential $\nu=0$. This is expected, as such mixing between
$\sigma$ and $\omega$ (or $\rho$-$a_0$) opens up new channels which
violate G-parity conservation. This can only happen when the ground 
state is not
invariant under charge conjugation as is the case in nuclear matter
with finite nucleon chemical potential.
In fact, 
isospin eigenstate $\omega$, which (ignoring small mixing with $\rho$ induced 
by the quark mass difference and electromagnetic effects ) in free space 
cannot decay into two pions, in a thermal bath
through its coupling with nucleon-antinucleon (or hole) excitation 
can decay into a pion pair with intermediate $\sigma$ meson 
\cite{wolf98}. The same argument
holds true for $\rho$ which in matter can decay
into $\pi - \eta$ \cite{teodorescu00}. 

Note that the mixing amplitude vanishes at
$|q|=0$ . Physically this corresponds to the situation when there is no way
for the vector particle to distinguish between longitudinal and transverse
direction and therefore rotational invariance forces them to be degenerate
forbidding mixing of different spin states in this limit. On the other hand,
it has also been observed that at very high momenta $\Pi_0$ falls off.
This corresponds to the situation when collective effects, which
basically related to the long wave length oscillations, die off with 
high momenta. This is a well-known phenomenon in many-body physics.

\subsection{Isoscalar channel : $\sigma$-$\omega$ mixing }
\begin{figure}
\begin{center}
\epsfig{file=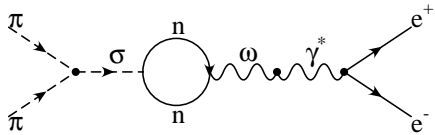,width=6.0cm,angle=0}
\end{center}
 \caption[Loop0]
  {\small The Feynman diagram for the process $\pi + \pi
\rightarrow e^+ + e^-$. \label{fig:loop0}}
\end{figure}

As has been mentioned before that $\sigma$-$\omega$ mixing has been studied
previously in the context of dilepton production for $\pi$-$\pi$ annihilation
in the s-wave channel in Ref.\cite{wolf98,saito98}. The thermal production
rates of the dilepton from this channel however not been estimated before.
In order to calculate the rate of dilepton production, we first calculate
the cross-section for the process $\pi + \pi \rightarrow \sigma
\rightarrow \omega \rightarrow e^+ + e^-$. The Feynman diagram for the 
relevant process are shown in fig \ref{fig:loop0}.

The interaction Lagrangian for the $\sigma$-$\pi\pi$ vertex is given by
\bea
{\cal L}=\frac{1}{2}g_{\sigma\pi\pi}m_\pi\phi_\sigma{\bf 
\vec{\phi}_\pi\cdot\vec{\phi}_\pi}
\eea

The cross-section can easily be evaluated to be :

\bea
\sigma_{\pi \pi \rightarrow e^+ e^-} &=& \frac{2 \pi \alpha^2}{3 q_z^2 M}
\frac {(g_{\sigma\pi\pi} m_\pi)^2}{g_\omega^2}\frac{F_\omega(M^2)G_\sigma(M^2)}
{{{\bf k}_\sigma}} 
{\mid\Pi_0\mid}^2 \ 
\nn
\eea
where, $\Pi_0$ is given by Eq.~(\ref{pi0}), and 
\begin{center}
\bea
F_\omega(M^2)=\frac{m_\omega^4}{(M^2 - m_\omega^2)^2 + m_\omega^2
\Gamma^2_\omega}\ ,
\nn \hspace*{10mm}
G_\sigma(M^2)=\frac{1}{(M^2 - m_\sigma^2)^2 + m_\sigma^2
\Gamma^2_\sigma(M)}\ ,
\nn
\eea
\end{center}

\bea
{\bf k}_\sigma = \frac{{\sqrt{(M^2-4 m_\pi^2)
}}}{2} \ .
\nn\eea

For $\omega$ we used a constant width of 9.5 MeV and for the sigma meson
we adopt the same mass and width as Ref.\cite{saito98}. 
Recent discussion on this
issue may be found also in Ref.\cite{tornquist96,harada96}. We also have
investigated the effects of form factors by considering the same 
form as in Ref. \cite{wolf98}: those are 
normalized to one for on-shell particles. Effects found are 
small as far as the dilepton yields are concerned. 

\subsection{Isovector channel : $\rho$-$a_0$ mixing }
 
\begin{figure}
\begin{center}
\epsfig{file=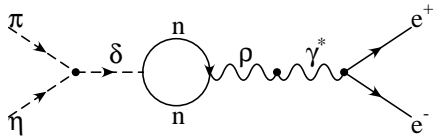,width=6.0cm,angle=0}
\end{center}
 \caption[Loop]
  {\small The Feynman  diagram for the process $\pi + \eta
\rightarrow e^+ + e^-$. \label{fig:loop}}
\end{figure}

Quite similar to $\sigma$-$\omega$ mixing $\rho$-$a_0$ mixing opens up
a new channel for dilepton production {\em viz.} 
$\pi + \eta \rightarrow e^+ + e^-$ in dense
nuclear matter through n-n excitations. The Feynman diagram of the
process is  shown in Fig.~\ref{fig:loop}.

To describe the $\pi a_0\eta$ vertex we use  
\bea
{\cal L}_{a_0\pi\eta}=f_{a_0\pi\eta}\frac{m_{a_0}^2-m_\eta^2}{m_\pi}
\phi_\eta{\vec \phi}_\pi\cdot{\vec\phi}_{a_0}\ .
\eea
For later convenience we define $g_{\pi a_0\eta}=f_{a_0\pi\eta}
(m_{a_0}^2-m_\eta^2)/m_\pi $. 

We should mention here that there is an uncertainty involved with the 
coupling parameter
$f_{a_0\pi\eta}$ as discussed in Refs.\cite{kirch96,pdg}. This arises
from the fact that $a_0$ lies close to the opening of the ${K{\bar K}}$ channel
leading to a cusp-like behavior in the resonant amplitude, therefore a naive
Breit-Wigner form for the decay width is inadequate.
Furthermore, as mentioned before, there is also uncertainty involved with
the $a_0$NN coupling which renders the precise extraction of $a_0 -
 \pi - \eta$ coupling even more difficult \cite{kirch96}.
We take  $f_{a_0\eta\pi}$=0.44 \cite{teodorescu00} which
gives $\Gamma_{a_0\rightarrow \pi \eta}$($m_{a_0}$)=$59$~MeV,
while the experimental vacuum width of $a_0$ is
between $50-100$~MeV \cite{pdg}. This choice is thus a conservative one.

As before, the cross-section involving mixing in the isovector channel
can be written in terms of momentum and temperature dependent mixing amplitude:

\bea
\sigma_{\pi \eta \rightarrow e^+ e^-} &=& \frac{2 \pi \alpha^2}{3 q_z^2 M}
\frac {g_{a_0\pi\eta}^2}{g_\rho^2}\frac{F_\rho(M^2)G_{a_0}(M^2)}
{{{\bf k}_{a_0}}} 
{\mid\Pi_0\mid}^2 \ 
\nn
\eea
where
\begin{center}
\bea
F_\rho(M^2)=\frac{m_\rho^4}{(M^2 - m_\rho^2)^2 + m_\rho^2
\Gamma^2_\rho(M)}
\nn \hspace*{10mm}
G_{a_0}(M^2)=\frac{1}{(M^2 - m_{a_0}^2)^2 + m_{a_0}^2
\Gamma^2_{a_0}(M)}
\nn
\eea
\end{center}

and the decay widths of the $a_0$ and $\rho$ mesons:
\bea
\Gamma_{a_0}(M)=
\frac{f_{a_0\pi\eta}^2}{8\pi}
(\frac{m_{a_0}^2-m_\eta^2}{m_\pi})^2
\frac{{\bf{ k}}_{a_0}}{M^2}\ , \hspace*{10mm}
\Gamma_\rho(M)=
\frac{g_{\rho\pi\pi}^2}{6\pi}
\frac{(\frac{M^2}{4}-m_\pi^2)^{\frac{3}{2}}
}{M^2}\ , 
\nn
\eea
\bea
{\bf k}_{a_0} = \frac{{\sqrt{(M^2-(m_\pi+m_\eta)^2)(M^2-(m_\pi-m_\eta)^2)
}} }{2 M}\ .
\nn\eea
As was the case for the $\omega$, we choose the mass of the $a_0$
according to Ref. \cite{pdg}.

\section{Dilepton production rates}
  
Once we know the relevant cross-section for the dilepton channel, the
thermal production rate of the lepton pairs
(number of reactions per unit time, per unit volume 
$R_{12}^{e^+e^-}=dN_{e^+e^- pairs}/d^4x$) can be estimated
in the independent particle approximation of kinetic theory \cite{gale87}:

\bea
\frac{dR_{12}^{e^+e^-}}{dM^2}=\int \frac{d^3k_1}{(2\pi)^3}f_1({\bf k}_1)
\int \frac{d^3k_2}{(2\pi)^3}f_2({\bf k}_2) 
\frac{d\sigma_{12}^{e^+e^-}}{dM^2}(s,M^2)v_{rel} 
\label{rate1}
\eea

where $f_1,f_2$ are the thermal distributions of the 1,2~species and
$v_{rel}=\frac{\lambda^{1/2}(s,m_1^2,m_2^2) \sqrt{s}}{2 E_1 E_2}$ with
$\lambda(x,y,z)=x^2+y^2+z^2-2xy-2xz-2yz$ is the triangle function.
This could be cast into the form:

\bea
\frac{dR_{12}^{e^+e^-}}{dM^2}=\frac{\lambda(M^2,m_1^2,m_2^2)}{2}\hspace*{2mm}
\Phi(s) 
\label{rate2}
\eea
\bea
\Phi(s)=\int \prod_{i=1,2} \frac{d^3p_i f_i(E_i)}
{(2\pi)^3E_i} \delta [s-(p_1+p_2)^2] \sigma_{12}^{e^+e^-}(E1,E2,M^2)
\label{phi1}
\eea

where $f_i (E_i)=\frac{1}{e^{\beta E_i}+1}$ are the distribution functions for $\pi$ and $\eta$ mesons assuming thermal equilibrium.
A departure from the standard notation should be noticed here. Usually, 
$\Phi(s)$ represents only the phase
space factor \cite{lichard94}. In our case the invariant mass 
cross-section depends also on energies through $\Pi_0$, we include that in
the definition of $\Phi(s)$.

After performing the angular integrations we get
\bea
\Phi(s)=\frac{1}{(2\pi)^4} \int_{m_1}^{\infty} dE_1 f_1 (E_1) 
\int_{E_2^-}^{E_2^+} dE_2 f_2 (E_2) \sigma_{12}^{e^+e^-}(E1,E2,M^2)
\label{phi2}
\eea

where $E_2^{\pm}=\frac{1}{2m_1^2}[E_1(s-m_1^2-m_2^2) \pm p_1 \lambda^{1/2}(s,m_1^2,m_2^2)] $. $\Phi (s)$ is then calculated numerically.

\subsection {Rates with $\sigma$-$\omega$ mixing}

\begin{figure} [htb!]
\begin{center}
\epsfig{file=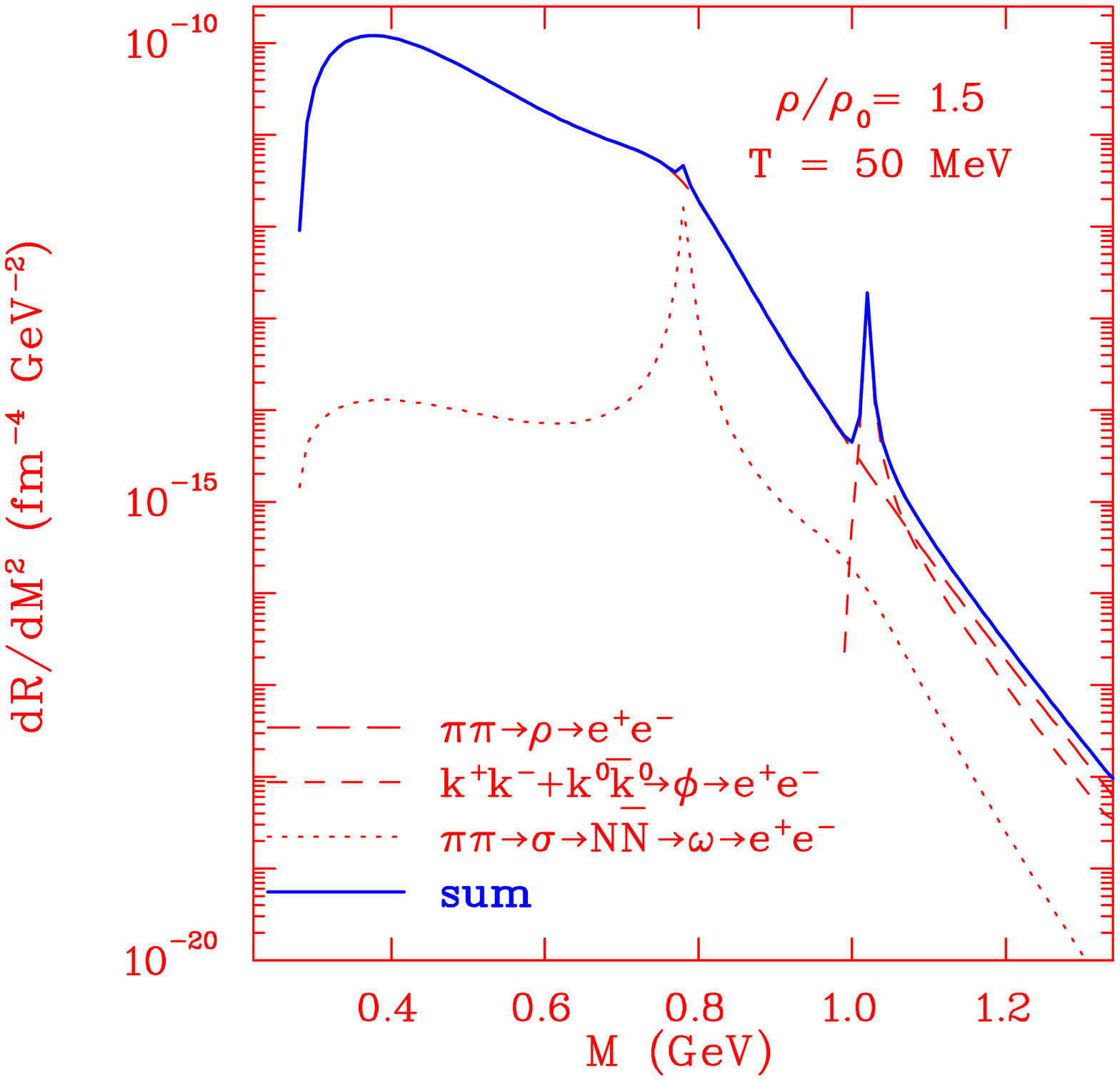,height=7cm,angle=0} 
\epsfig{file=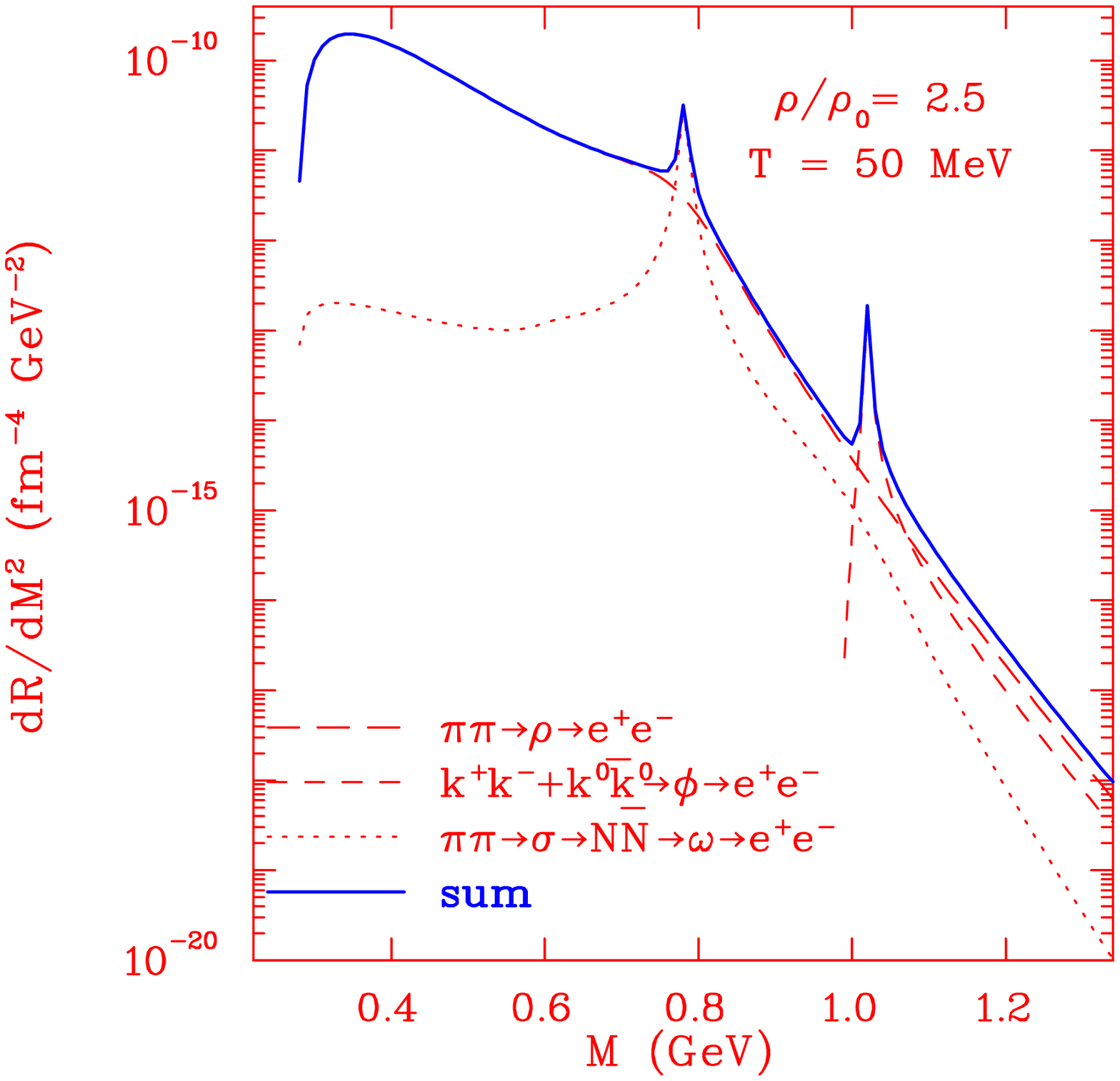,height=7cm,angle=0} 
\end{center}
 \caption[os50] 
  {\small Effect of $\sigma$-$\omega$ mixing on the dilepton production rate
at T = 50 MeV for two different densities as mentioned in the legend.
\label{fig:os50}}
\end{figure}

First we concentrate on the isoscalar channel or in other words the
effect of $\sigma$-$\omega$ mixing on the dilepton production rate. As the
mixing amplitude now involves the $\omega$ propagator we observe a sharp 
peak in the vicinity of the  $\omega$ mass. Obviously this mixing
amplitude depends on the density and the temperature. The effects are
more pronounced in Fig.~\ref{fig:os50}
where we have higher density. No such peak
is observed for the $\sigma$ meson which is sufficiently broad. 
We also plot the rates for $\pi$-$\pi$ and $K$-${\bar K}$ annihilation for
comparison. The left panel shows the results at $\rho/\rho_0=1.5$ and the
right panel shows the same for $\rho/\rho_0=2.5$. 

\begin{figure} [htb!]
\begin{center}
\epsfig{file=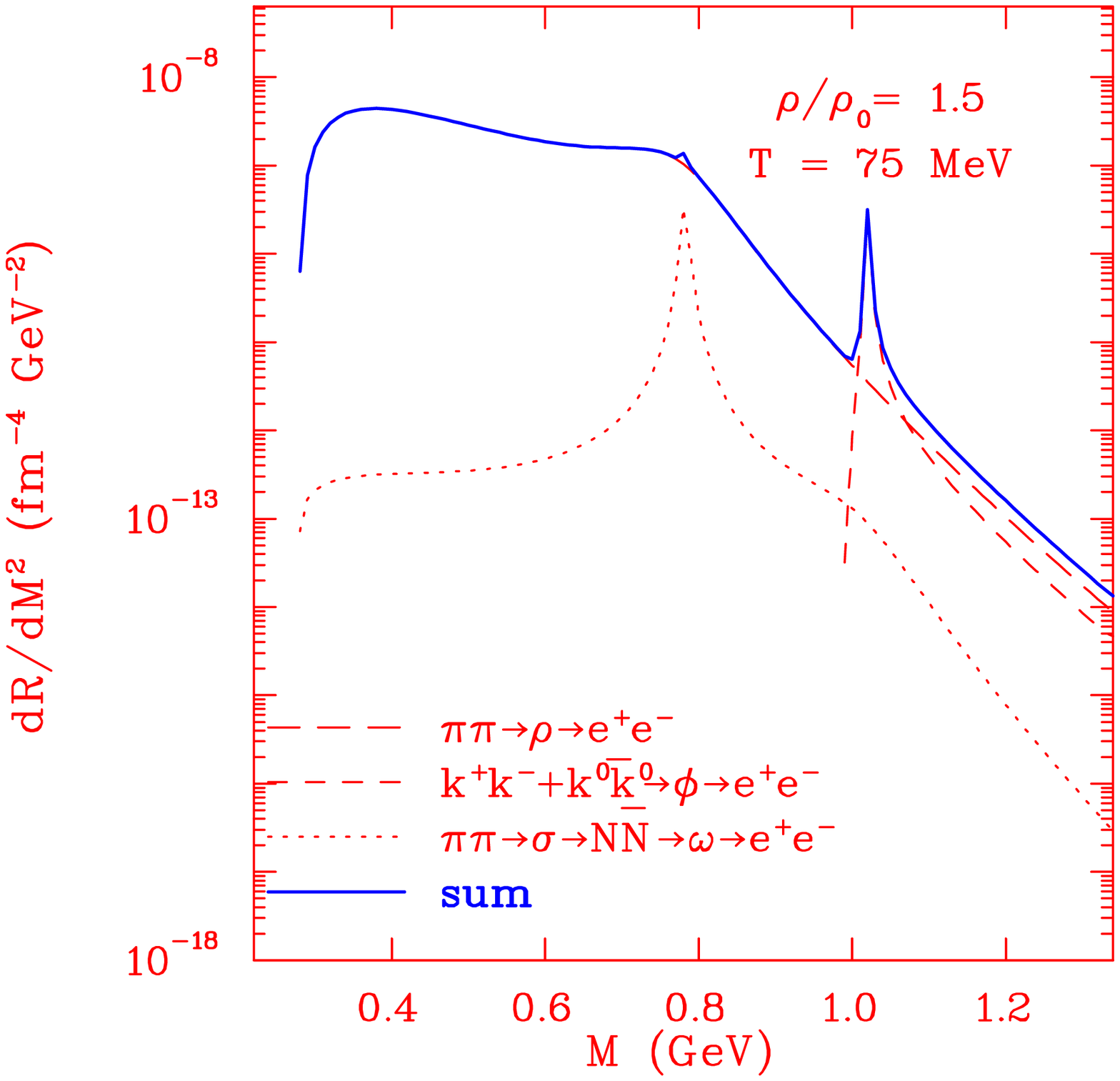,height=7cm,angle=0} 
\epsfig{file=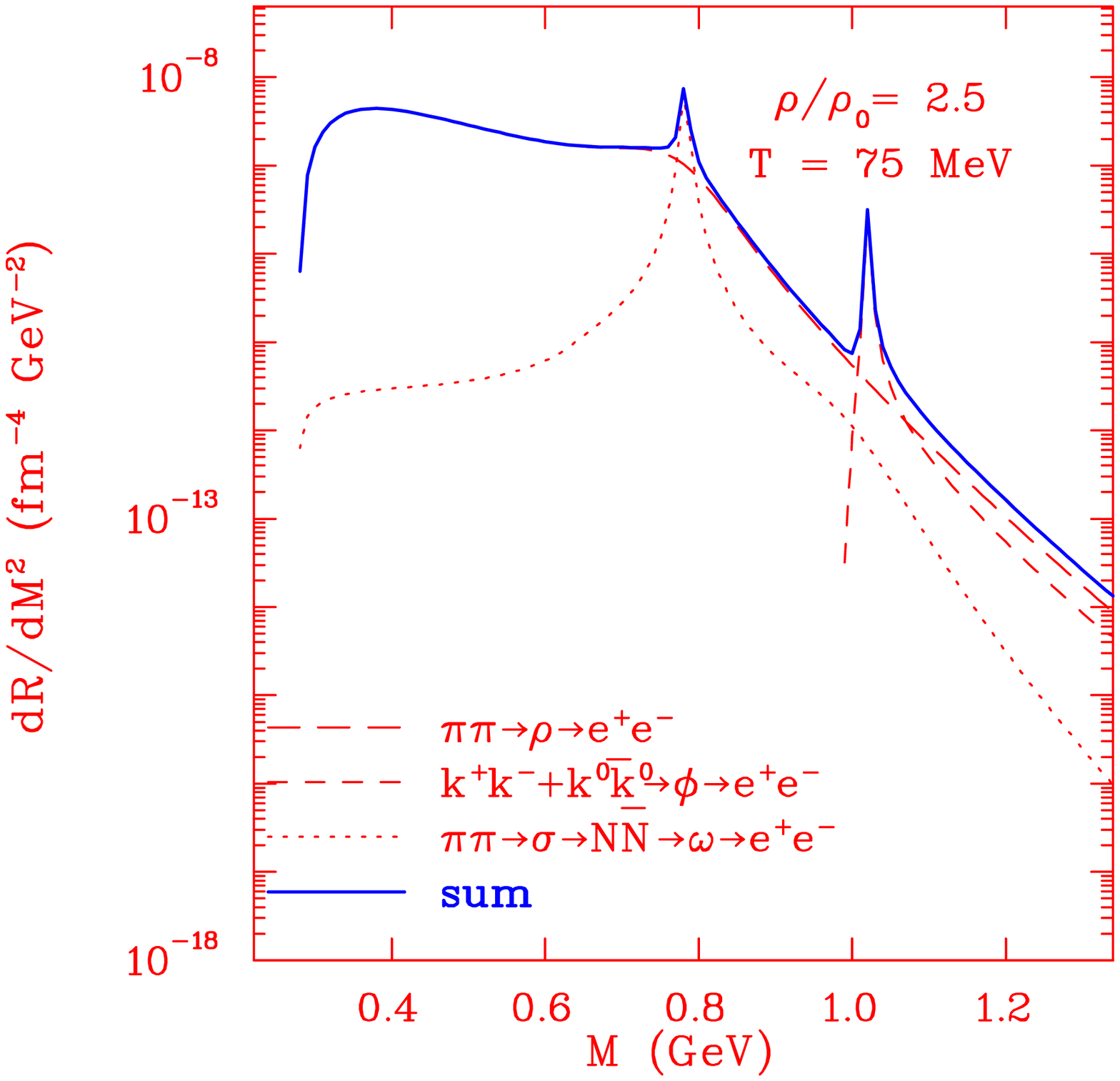,height=7cm,angle=0} 
\end{center}
 \caption[os75] 
  {\small Same as Fig.~\ref{fig:os50} but for T=75 MeV\label{fig:os75}}
\end{figure}

Fig.\ref{fig:os75} shows the results for the isoscalar channel at a 
higher temperature. It is evident that at higher density the effect of mixing
is also high. 

It is to be noted that to calculate the dilepton rate we have used tree
level propagator for $\omega$ and $\sigma$. 
In Ref.\cite{wolf98} it is argued that $\sigma$-$\omega$ mixing broadens
the $\omega$ decay width substantially in nuclear matter at high density.
This is understandable as now $\omega$ thorough nucleon-antinucleon (hole)
excitations can decay into pion pairs as opposed to
vacuum because of G-parity violation in matter. Therefore in matter this
decay channel enjoys a larger phase space and becomes as broad as $\rho$
as its vacuum mass is almost degenerate with that of $\rho$ meson. But
at the one loop level the diagonal elements of the mixed propagator would
also reduce its mass (off-diagonal elements have little effect on the 
effective mass). Now if one considers the broadening induced by the 
off-diagonal element of the dressed propagator of the $\omega$ meson, then
to be consistent one should also take the medium modified mass for the $\omega$
meson. It is well-known that in Walecka model $\omega$ meson mass also
drops \cite{hatsuda} which would again reduce the phase space.  These two 
competing effects are not expected therefore to change the in-medium
width of $\omega$ significantly. We, however, adopt a different point of 
view.  Our focus here is to see the effect of mixing on the 
dilepton production 
and how it compares with $\pi$-$\pi$ annihilation. 
It should be noted that the contribution from $\sigma$-$\omega$ mixing 
does not change the resultant
spectra much expect a small peak representing $\omega$ meson pole in the
transition amplitude. However, we expect an $\omega$ peak in the dilepton
spectra even in the absence of meson mixing. The situation, as we shall see,
is different for the isovector mode. There the dominant contributions are 
distributed near the tail 
of the $\pi$-$\pi$ annihilation spectra ($\rho$ spectral
function). On the other hand if $\rho$ spectral function gets sufficiently
broadened this effect might become more visible near the $\omega$ peak region
where additional contributions would come from the mixing. 

\subsection {Rates with $\rho$-$a_0$ mixing}
\begin{figure} [htb!]
\begin{center}
\epsfig{file=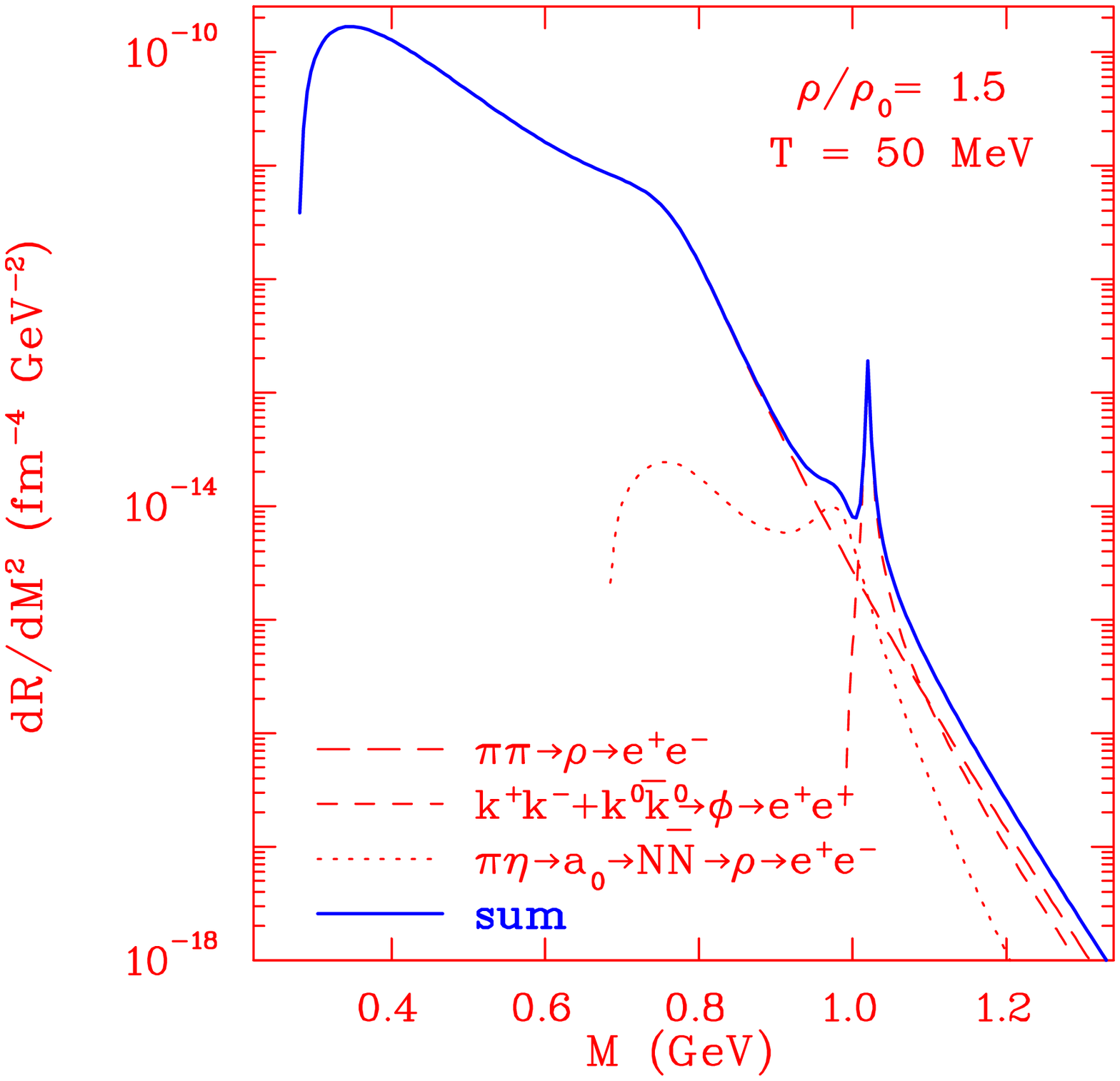,height=7cm,angle=0}
\epsfig{file=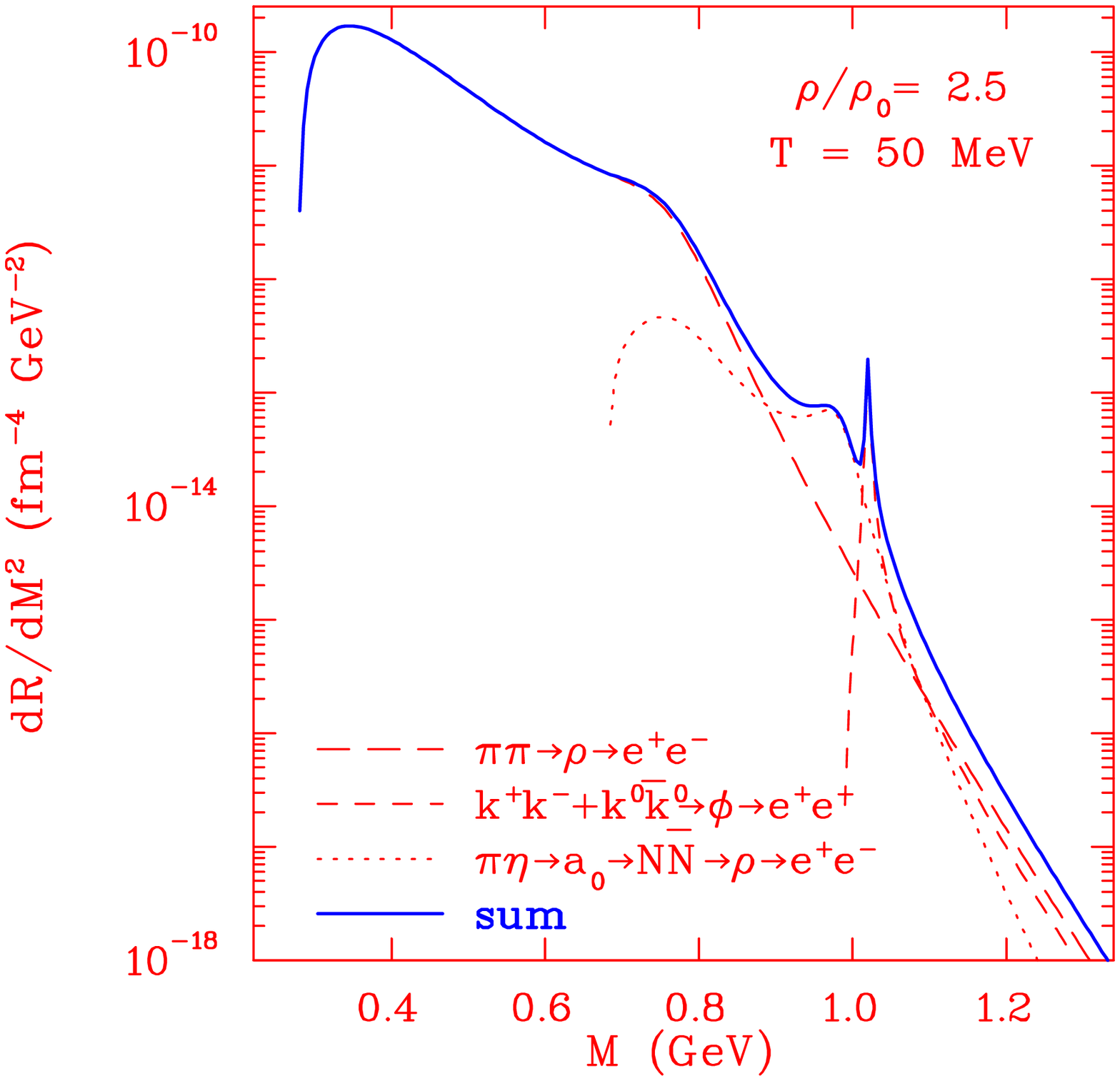,height=7cm,angle=0}
\end{center}
 \caption[rd50]
  {\small Effect of mixing on dilepton production rate in the isovector
channel involving $\rho$-$a_0$ mixing at T=50 MeV.
\label{fig:rd50}}
\end{figure}

\begin{figure} [htb!]
\begin{center}
\epsfig{file=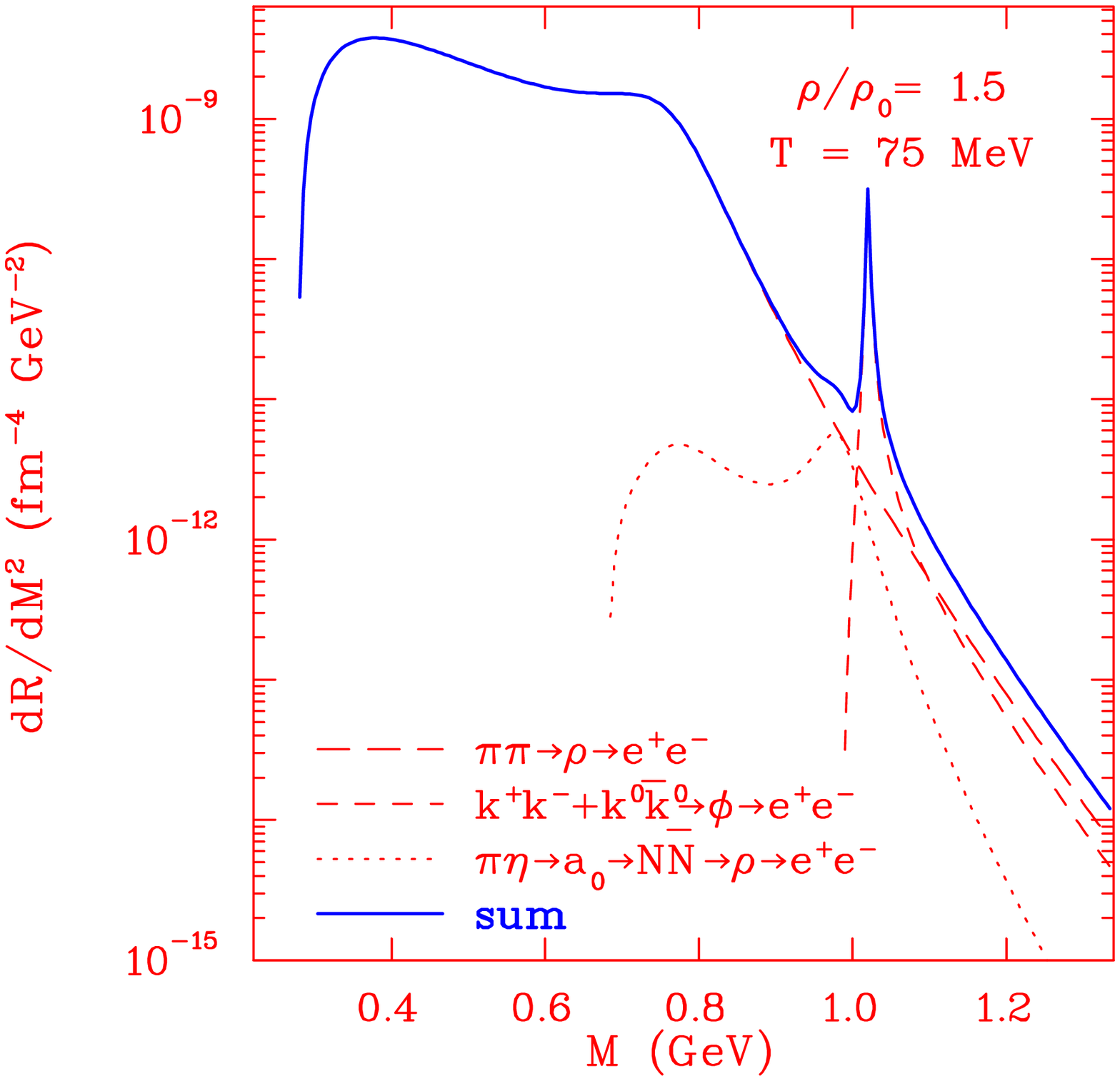,height=7cm,angle=0}
\epsfig{file=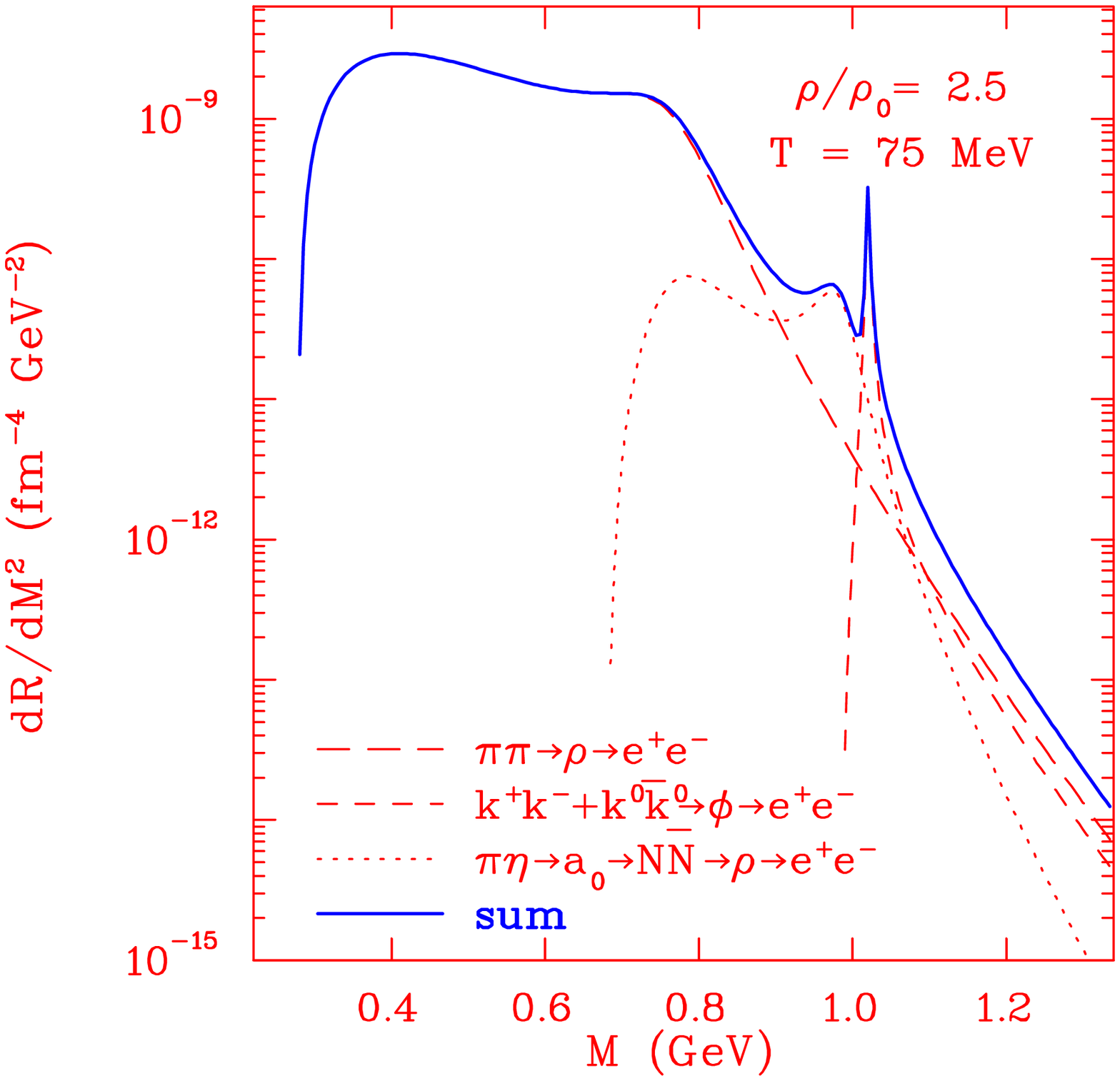,height=7cm,angle=0}
\end{center}
 \caption[rd75]
  {\small Same as Fig.~\ref{fig:rd50} but for T=75 MeV\label{fig:rd75}}
\end{figure}

Next we discuss the effect of mixing in the isovector channel which involves
$\rho$-$a_0$ mixing in Fig.\ref{fig:rd50} (\ref{fig:rd75}) for T=50 MeV (75 MeV). This induces an
$a_0$ bump 
just below the $\phi$  peak. $\rho$-$a_0$ channel seems to be more 
interesting as it induces additional contribution beyond the tail of $\rho$
spectral function.  One can easily observe from Fig \ref{fig:rd50}
that even at density $\rho/\rho_0=1.5$ the contribution of the mixing 
wins over the $\pi$-$\pi$ annihilation rate in the vicinity of M = 1.0 GeV. 
Naturally at higher density this goes up as evident from the right panel
of Fig. \ref{fig:rd50}. 

Interestingly enough the dilepton yield induced by the mixing in the isovector
channel at higher temperature also seems to be significant as evident from
the figures below.

Note that, unlike $\omega$, the broadening
induced by the mixing on $\rho$ decay width has been found to be 
marginal. One can understand this physically. Mixing opens up a new channel 
for $\rho$ to decay into $\pi$-$\eta$. $\eta$ being quite heavy only the 
tail part of the $\rho$ can contribute. Therefore it does not enjoy much of 
phase space to decay into $\pi-\eta$. 
It does seem reasonable to conclude that $\rho$-$a_0$ mixing provides
more promising kinematical window in the invariant mass spectra to study
exclusive in-medium effects.

\subsection {Rates with mixing both in isoscalar and isovector channels}

\begin{figure} [htb!]
\begin{center}
\epsfig{file=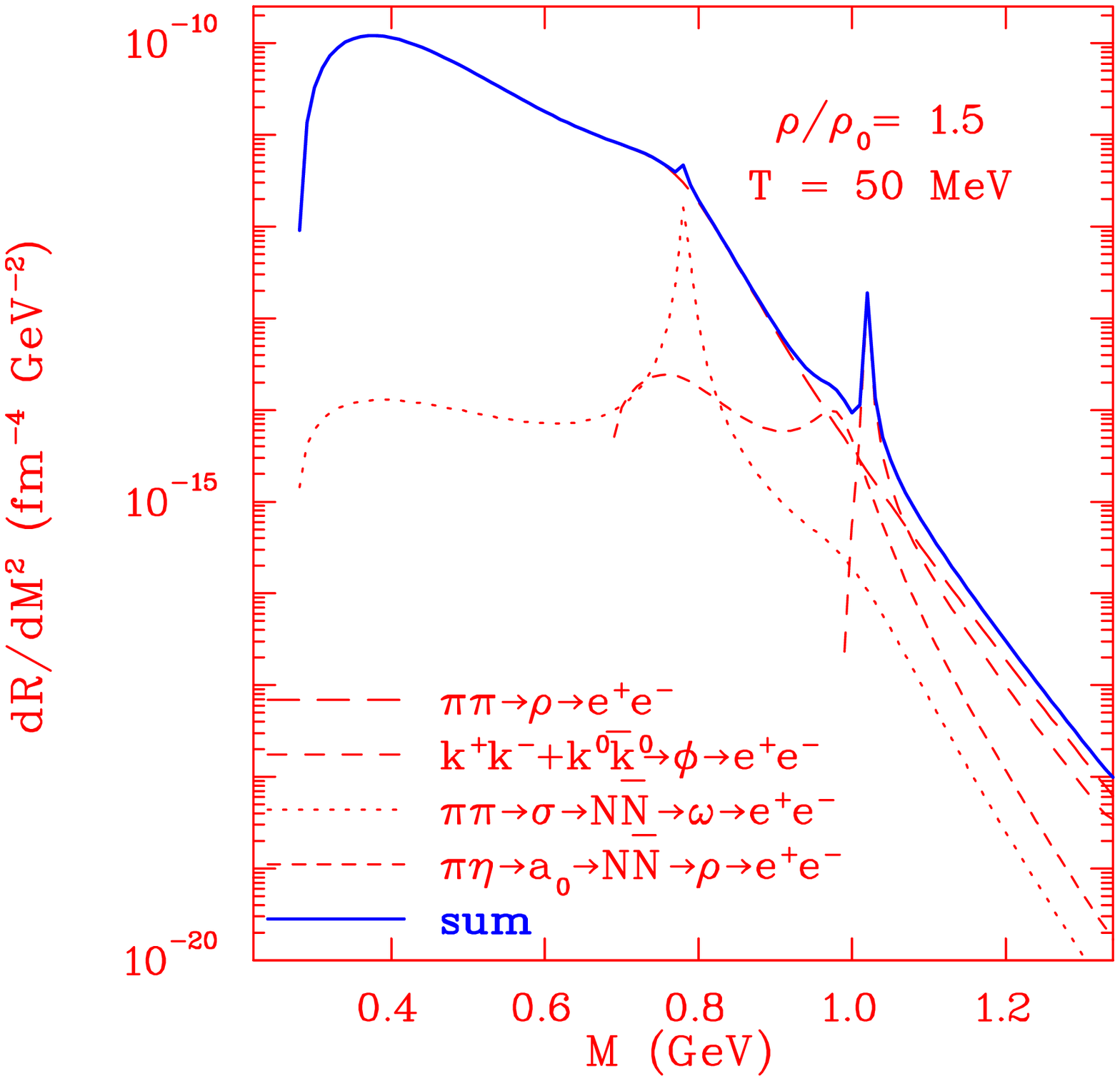,height=7cm,angle=0} 
\epsfig{file=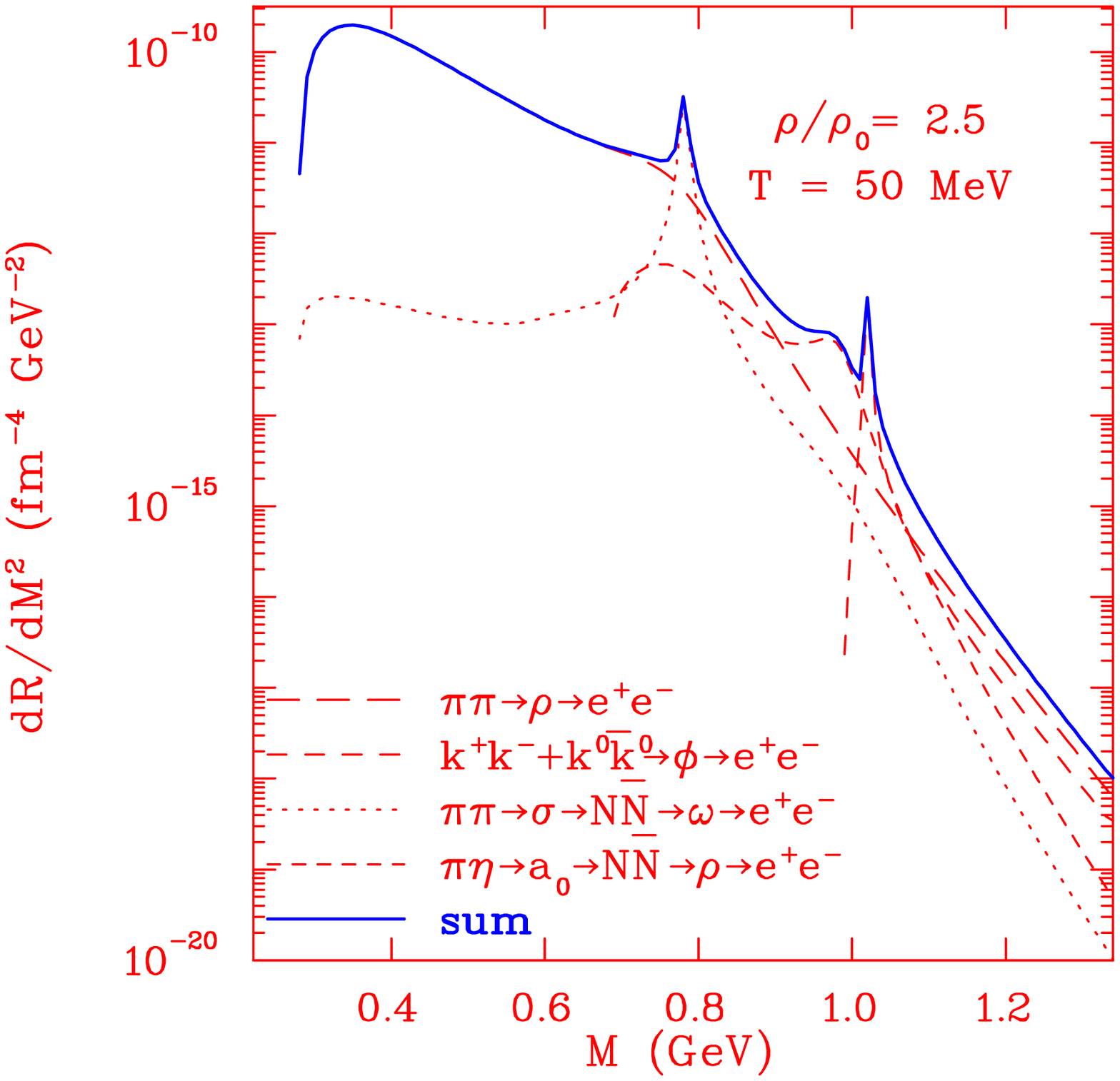,height=7cm,angle=0} 
\end{center}
 \caption[spectra2550] 
  {\small Rates at T = 50 MeV with mixing in both isoscalar and isovector
channels.
\label{fig:spectra50}}
\end{figure}

\begin{figure} [htb!]
\begin{center}
\epsfig{file=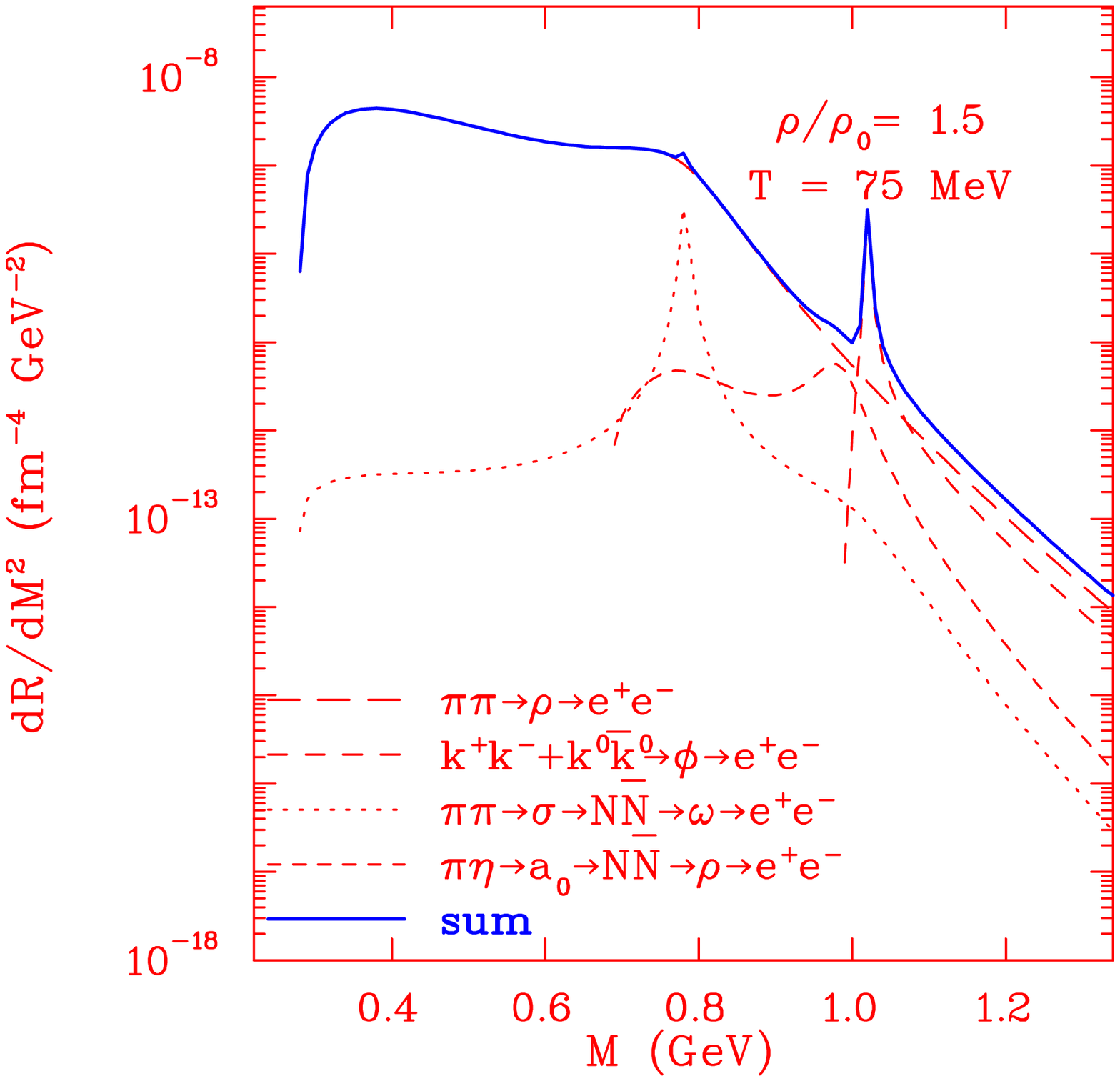,height=7cm,angle=0} 
\epsfig{file=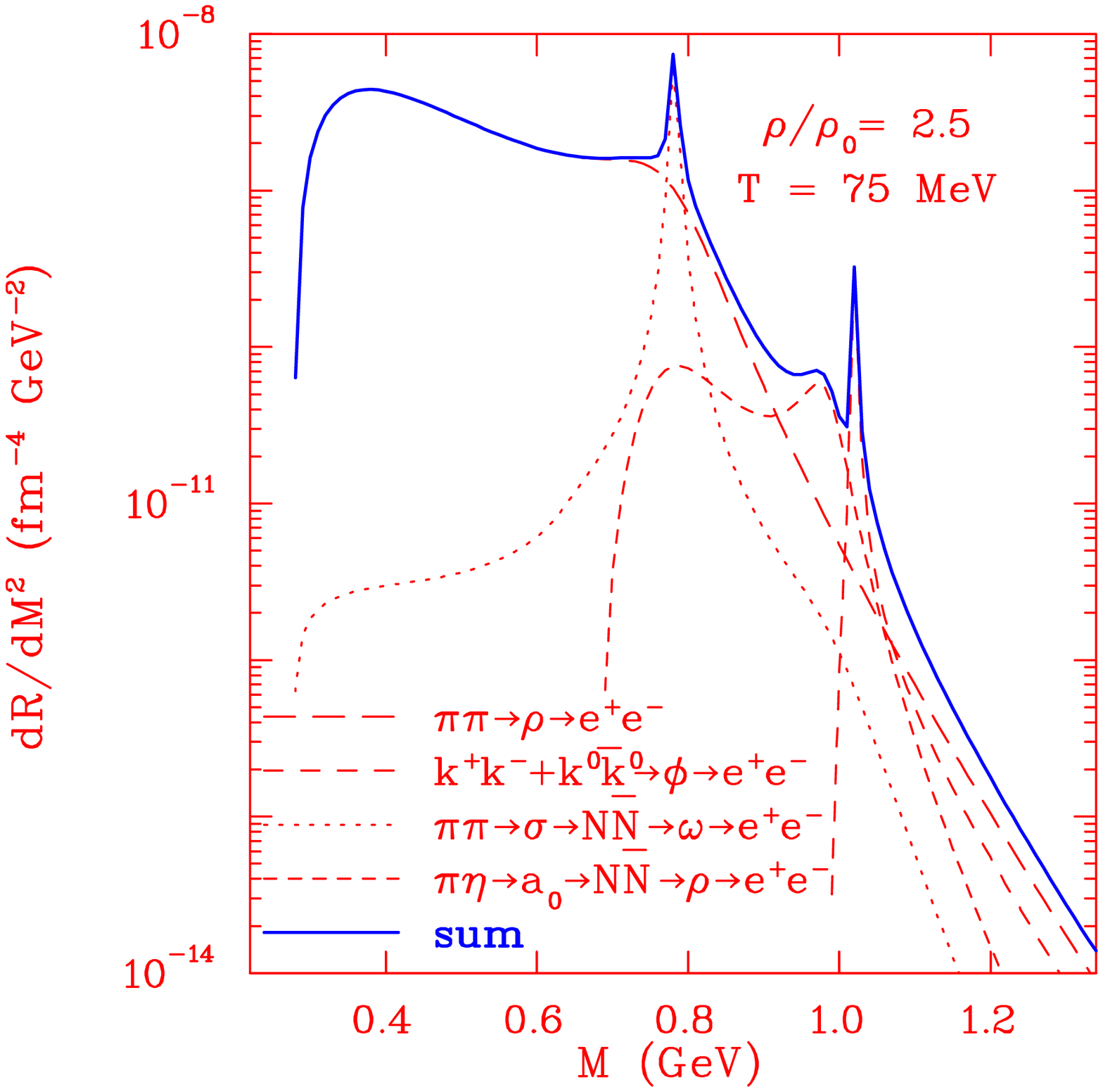,height=7cm,angle=0} 
\end{center}
 \caption[spectra2575] 
  {\small Rates at T = 75 MeV with mixing in both isoscalar and isovector
channels.
\label{fig:spectra75}}
\end{figure}

For completeness we show in Figs.~\ref{fig:spectra50}~and~\ref{fig:spectra75} 
the total dilepton yield with the combined effects
of mixing in both the isoscalar and isovector channels. It is 
important to see that rates induced by mixing are significant
in certain windows of density and temperature.

\section{ Summary and Conclusion}

 To conclude and summarise, we have estimated the
dilepton production rates arising from  scalar and vector meson
mixing in different isospin channel involving nucleon
-hole (or anti-nucleon) excitations as intermediate states.
The rates have been compared against the $\pi$-$\pi$ annihilation
in the p-wave channel ($\rho$-channel) and $K$-${\bar K}$
annihilation. Higher yield of dileptons in the invariant mass region
between the $\rho$ and the $\phi$ peaks are expected. The mixing 
in the isoscalar
channel contributes mostly between the two-pion threshold and the 
$\omega$ peak but is small compared to the 
$\pi$-$\pi$ annihilation
contribution. On the other hand, the isovector channel seems to provide a
better probe as it contributes near the tail region of the two pion
annihilation contribution. It should be stressed here that broadening or
dropping $\rho$ meson mass would even favor our observation. In that
case the $\pi$-$\pi$ background would be pushed more towards lower
invariant mass region bringing the $a_0$ peak into a clearer relief. 
Finally, it is important to realize that mixing
effects only appear in the longitudinal part of the vector meson 
spectral function and therefore there should be no effect of the mixing on 
the photon spectra which involves the transverse part of the vector meson 
polarization tensor.

The effects studied here are density driven and can 
therefore be observed in environments where one has high baryonic chemical
potential and comparatively lower temperature. The most relevant experiments
to observe this kind of medium-induced phenomenon are the ones to be performed
at GSI using the HADES detector where the density might
reach up to three times the normal nuclear matter density and temperature 
T=60-90 MeV \cite{friese99}. Ref.\cite{devismes00} also indicates a similar 
range of densities and temperatures at SIS/GSI energies. Those
observations have guided our choice of the temperatures and densities used
in this study.

The calculations performed here can, and will be improved upon. It is
important to proceed cautiously. As a
next step, it would be necessary to address this problem with 
the full machinery of finite temperature 
many-body theory. This means, among other things, the 
dressing of the propagators and the consistent inclusion of baryonic
resonances. Also, the interaction zone
in a heavy collision is but a transient medium and the dynamics will
contribute to mask our signal to some extent. However, owing to the
nature of the electromagnetic signal, radiation from the dense
regions will emerge. In our approach those dynamics would impact on
the height of the $a_0$ peak. Similarly,
a realistic assessment of the initial  states of the
dilepton-producing channels should involve a realistic simulation. A
transport treatment can therefore address the concerns raised above. 
Such investigations are under progress.  

\acknowledgements

This work was supported in part by the Natural Sciences and Engineering
Research Council of Canada and in part by the Fonds FCAR of the
Qu\'ebec Government.


\end{document}